\documentstyle[epsf]{elsart}                                     

\def\kr78{$^{78}$Kr}
\def\bra{\langle} \def\ket{\rangle} 
\begin{document}
\begin{frontmatter}

\title{Projected shell model study for the yrast-band structure 
of the proton-rich mass-80 nuclei}
\author{R. Palit$^{(1)}$, J.A. Sheikh$^{(1,2)}$, Y. Sun$^{(3,4)}$ and H.C. Jain$^{(1)}$}
\address{$^{(1)}$Tata Institute of Fundamental Research, Mumbai 400 005, India
\\
$^{(2)}$Physik-Department, Technische Universit\"at M\"unchen,
D-85747 Garching, Germany\\
$^{(3)}$Department of Physics and Astronomy, University of Tennessee,
Knoxville, Tennessee 37996, U.S.A. \\
$^{(4)}$Department of Physics, Xuzhou Normal University,
Xuzhou, Jiangsu 221009, P.R. China}
\maketitle

\begin{abstract}
A systematic study of the yrast-band structure for the 
proton-rich, even-even mass-80 nuclei is carried out
using the projected shell model approach. 
We describe the energy spectra, 
transition quadrupole moments and gyromagnetic factors. The
observed variations in energy spectra and transition quadrupole moments 
in this mass region are
discussed in terms of the configuration mixing of the projected deformed 
Nilsson states as a function of shell filling.
\end{abstract}
\end{frontmatter}

\section{Introduction}
The study of low- and high-spin phenomena in the proton-rich mass-80 nuclei have attracted considerable 
interest in recent years. This has been motivated by the increasing power of experimental facilities and
improved theoretical descriptions, as well as by the astrophysical requirement in understanding
the structure of these unstable nuclei.
In comparison to the rare-earth region where the change  
in nuclear structure properties is quite smooth with respect to particle number,
the structure of the proton-rich mass-80 nuclei shows considerable variations when going from one
nucleus to another. 
This is mainly due to the fact that the available shell model configuration 
space in the mass-80 region is much smaller than in the rare-earth region. 
The low single particle level density implies that a drastic change near the Fermi surfaces can occur
among neighboring nuclei. Another fact is that in these medium mass proton-rich nuclei,
neutrons and protons occupy the same single particle orbits.  
As the nucleus rotates, 
pair alignments of neutrons and protons compete with each other and
in certain circumstances they can align simultaneously. 

Thanks to the new experimental facilities, in particular, to the newly constructed detector arrays,
the domain of nuclides accessible for spectroscopic studies has increased drastically
during the past decade. For example, some extensive
measurements \cite{sch96,paul97,gr89,gp99,ch97,heese91,algora00,ru97,kr72,li87} of the 
transition quadrupole-moments, extracted from the 
level lifetimes and excitation energy of 2$^+$ state, have been carried out for 
Kr-, Sr- and Zr-isotopes. These measurements have revealed large variations in nuclear structure 
of these isotopes with respect to particle number and angular momentum. It has been
shown that alignment of proton- and neutron-pairs at higher angular momenta can change the nuclear
shape from prolate to triaxial and to oblate.
      
A systematic description for all these observations 
poses a great challenge to theoretical models. 
The early mean-field approaches were devoted to a general study of the
structure in the mass-80 region. Besides the work 
using the Woods-Saxon approach \cite{na85}, there were studies using 
the Nilsson model \cite{he84} and the Skyrme Hartree-Fock + BCS theory \cite{bo85}. 
More recently, microscopic calculations were performed using the 
Excited VAMPIR approach \cite {petrovici96,petrovici00}. However, this approach is numerically
quite involved and is usually employed to study some quantities for selected nuclei.
The large-scale spherical shell model diagonalization calculations 
\cite{cau94} have  been recently successful
in describing the $pf$ shell nuclei, but the configuration space required for studying the
well-deformed mass-80 nuclei is far beyond what the modern computers
can handle. 
 
Recently, the projected shell model (PSM) \cite{review} has become quite popular to study the
structure of deformed nuclei. The advantage in this method is that the
numerical requirement is minimal and, therefore, it is possible to perform a systematic study for
a group of nuclei in a reasonable time frame. The PSM approach is based on the
diagonalization in the angular-momentum
projected basis from the deformed Nilsson states. A systematic study of the rare-earth 
nuclei \cite {hara91,ysun94} has been carried out and the agreement 
between the PSM results and experimental data has been found to be quite good.
Very recently, the PSM approach has also been used to study the high-spin properties
of $^{74}$Se \cite{do98}, which lies in the mass-80 region.

The purpose of the present work is to perform a systematic PSM study of the low- and high-spin 
properties for the proton-rich, even-even Kr-, Sr- and Zr-isotopes. 
The physical quantities to be described are energy spectrum,
transition quadrupole moment and gyromagnetic factor. 
In addition, to compare the avaliable data with theory in a systematic way, 
we also make predictions for the structure of the $N=Z$ nuclei which could  
be tested by future experiments with radioactive ion beams.
We shall begin our discussion with an outline of the PSM in section II. 
The results of calculations and comparisons with experimental data are presented in section III.
Finally, the conclusions are given in section IV.

\section{The Projected shell model}

In this section, we shall briefly outline the basic philosophy of the PSM. 
For more details about the model, the reader is referred to the review 
article \cite{review}. The PSM 
is based on the spherical shell model concept. It differs from the
conventional shell model in that
the PSM uses the angular momentum projected states as
the basis for the diagonalization of the shell model Hamiltonian. 
What one gains by starting from a deformed basis 
is not only that shell model calculations for heavy nuclei
become feasible but also physical interpretation for the complex systems
becomes easier and clearer. 

The wave function in the PSM is given by
\begin{equation}
|\sigma,IM \ket = \sum_{K,\kappa}f^{\sigma}_{\kappa}\hat{P}^{I}_{MK}|\phi_\kappa \ket .
\end{equation} 
The index $\sigma$ labels the states with same angular momentum and $\kappa$ 
the basis states. $ \hat{P}^{I}_{MK} $ is angular momentum projection operator 
and $f^{\sigma}_{\kappa} $ are the weights of the basis states $\kappa$.  
 
We have assumed axial-symmetry for the basis states and the intrinsic states are, 
therefore, the eigenstates of the $K$-quantum number.
For calculations of an even-even system, the following four kinds of basis states 
$|\phi_\kappa\ket$ are considered:
the quasiparticle (qp) vacuum $| 0 \ket$, two-quasineutron states
${a_{\nu_1}^\dagger} {a_{\nu_2}^\dagger} | 0 \ket$, two-quasiproton states    
${a_{\pi_1}^\dagger} {a_{\pi_2}^\dagger} | 0 \ket$, and two-quasineutron plus
two-quasiproton (or 4-qp) states
${a_{\nu_1}^\dagger}{a_{\nu_2}^\dagger}{a_{\pi_1}^\dagger}{a_{\pi_2}^\dagger}|0\ket$.
The projected vacuum $| 0 \ket$, for instance, is the ground-state band (g-band)
of an even-even nucleus. 

The weight factors, $f^{\sigma}_\kappa$ in Eq. (1), are determined by diagonalization of
the shell model Hamiltonian in the space spanned by the projected basis states 
given above. This leads to
the eigenvalue equation
\begin{equation}
\sum_{\kappa '} (H_{\kappa \kappa'} - E_\sigma N_{\kappa \kappa'} ) f^{\sigma}_{\kappa'} = 0 ,  
\end{equation}
and the normalization is chosen such that 
\begin{equation}
\sum_{\kappa \kappa'}f^\sigma_\kappa N_{\kappa \kappa'} f^{\sigma'}_{\kappa'}
= \delta_{\sigma \sigma'},
\end{equation}
where the Hamiltonian and norm matrix-elements are given by 
\begin{eqnarray}
H_{\kappa \kappa'} & = & \bra \phi_\kappa | \hat {H} \hat{P}^{I}_{K_\kappa 
K'_{\kappa'}} | \phi_{\kappa'} \ket ,\\
N_{\kappa \kappa'} &  = & \bra \phi_\kappa | \hat{P}^{I}_{K_\kappa K'_{\kappa'}} | \phi_{\kappa'} \ket .
\end{eqnarray}
In the numerical calculations, we have used the standard quadrupole-quadrupole
plus (monopole and quadrupole) pairing force, i.e.
\begin{equation}
\hat{H} = \hat{H_0} - \frac{1}{2} \chi \sum_\mu {\hat{Q}_\mu}^\dagger 
{\hat{Q}_\mu} - G_M \hat{P}^\dagger \hat{P} - G_Q \sum_\mu {\hat{P}_\mu}^\dagger \hat{P}_\mu ,
\end{equation}
where $\hat{H_0}$ is the spherical single-particle Hamiltonian. The strength of the 
quadrupole force $\chi$ is adjusted such that the known
quadrupole deformation parameter $ \epsilon_2 $ is obtained. This condition results from the 
mean-field approximation of the quadrupole-quadrupole interaction of the Hamiltonian in
Eq. (6). The monopole pairing force constants $G_M$ are adjusted to give 
the known energy gaps. For all the calculations in this paper, we have used \cite{do98}  
\begin{eqnarray}
G_M ^\nu & = & \left[ 20.25 - 16.20 \frac{N-Z}{A} \right] A^{-1} , \nonumber \\
G_M ^\pi & = &   20.25 A^{-1} . 
\end{eqnarray}
The strength parameter $G_Q$ for quadrupole pairing is assumed to be 
proportional to $G_M$. It has been shown that the band crossing spins 
vary with the magnitude of the quadrupole pairing force \cite{Sun94}. 
The ratio $G_Q/G_M$ has therefore been slightly adjusted around 0.20
to reproduce the observed band crossings.  

Electromagnetic transitions can give important 
information on the nuclear structure and provide a stringent test
of a particular model. In the present work, we have calculated the 
electromagnetic properties
using the PSM approach.
The reduced transition probabilities $B(EL)$ from the initial state 
$( \sigma_i , I_i) $ to the final state $(\sigma_f, I_f)$ are given by 
\cite {ysun94}
\begin{equation}
B(EL,I_i \rightarrow I_f) = {\frac {e^2} {(2 I_i + 1)}} 
| \bra \sigma_f , I_f || \hat Q_L || \sigma_i , I_i \ket |^2 , 
\end{equation}
where the reduced matrix-element is given by
\begin{eqnarray*}
\bra \sigma_f , I_f || \hat Q_L || \sigma_i , I_i \ket  &  = &
\sum_{\kappa_i , \kappa_f} f_{\kappa_i}^{\sigma_i} f_{\kappa_f}^{\sigma_f}
\sum_{M_i , M_f , M} (-)^{I_f - M_f}  \left( \begin{array}{ccc}
I_f & L & I_i \\
-M_f & M & M_i 
\end{array} \right) \nonumber \\
 & & \times \bra \phi_{\kappa_f} | {\hat{P}^{I_f}}_{K_{\kappa_f} M_f} \hat Q_{LM}
\hat{P}^{I_i}_{K_{\kappa_i} M_i} | \phi_{\kappa_i} \ket \nonumber \\
\end{eqnarray*}
\begin{eqnarray}
 =& 2\sum_{\kappa_i , \kappa_f} f_{\kappa_i}^{\sigma_i} f_{\kappa_f}^{\sigma_f}
\sum_{M^\prime,M^{\prime\prime}} (-)^{I_f-K_\kappa f} (2 I_f + 1)^{-1}
\left( \begin{array}{ccc}
I_f & L & I_i \\
-K_{\kappa_f} & M^\prime & M^{\prime\prime}
\end{array} \right)  \nonumber \\
 & \times \int d\Omega D_{M'' K_{\kappa_i}} (\Omega) 
\bra \phi_{\kappa_f} | \hat Q_{LM'} \hat{R}(\Omega) | \phi_{\kappa_i} \ket .  
\end{eqnarray}
The transition quadrupole moment $Q_t(I) $ is related to the $B(E2)$ 
transition probability through
\begin{equation}
Q_t (I) = \left( \frac{16 \pi}{5} \frac{ B(E2,I\rightarrow I-2)} {\bra I K 2 0 | I-2 K \ket} \right)^{1/2} .
\end{equation}
In the calculations, we have used the effective charges of 1.5e for protons 
and 0.5e for neutrons, which are the same as in the previous PSM calculations \cite{review,ysun94}. 

Variation of $Q_t$ with spin $I$ can provide information on shape evolution in nuclei.
In the case of a rigid rotor, the $Q_t$ curve as a function of $I$ is a 
straight line. Experimentally, 
one finds a deviation from the rigid body behavior for most of the nuclei in this mass region. One
expects more or less a constant value in $Q_t$ up to the first band crossing. At
the band crossing, one often sees a dip in $Q_t$ value due to a small overlap
between the wave functions of the initial and the final states involved. There
can be a change in $Q_t$ values after the first band crossing, which indicates a
shape change induced by quasiparticle alignment. 

The other important electromagnetic quantity, which can give crucial information
about the level occupancy and thus is an direct indication of the nature of alignment, is 
the gyromagnetic factor (g-factor). 
The g-factors $ g(\sigma, I), g_p(\sigma,I) $ and $ g_n(\sigma, I) $
are defined by \cite {ysun94}
\begin{equation}
g(\sigma, I) = \frac {\mu(\sigma, I)}{\mu_N I} = g_\pi (\sigma,I) + g_\nu (\sigma,I) , 
\end{equation}
with $g_\tau (\sigma, I), \tau = \pi, \nu $ given by
\begin{eqnarray}
g_\tau (\sigma, I) & = & \frac {1}{ \mu_N [I(I+1)]^{1/2} } \nonumber  
\end{eqnarray}
\begin{eqnarray}
  \times \left[ g_l^\tau < \sigma, I || \hat J^\tau || \sigma, I > 
+ (g^\tau_s - g^\tau_l) \bra \sigma,I || \hat S^\tau || \sigma, I \ket \right] .
\end{eqnarray}
In our calculations, the following standard values of $g_l $ and $g_s$ have been
taken: 
$ g_l^\pi = 1, g_l^\nu = 0, g_s^\pi = 5.586 \times 0.75 $ and $ g_s^\nu = -3.826 \times 0.75 $.
Unfortunately, very few experiments \cite
{bi93,ku89,te99} have been done to measure the g-factors in this mass region.
The only experiment which has recently been performed is for $^{84}$Zr \cite{te99}.
In the present paper, we have calculated g-factors for this nucleus and compared
with the data. 

\section{Results and Discussions}

The PSM calculations proceed in two steps. In the first step, an optimum set of
deformed basis is constructed from the standard Nilsson potential. The Nilsson parameters
are taken from Ref. \cite{be85} and the calculations are performed by considering
three major shells (N = 2, 3 and 4) for both protons and neutrons. 
The basis deformation $\epsilon_2 $ used for each nucleus is taken  
either from experiment if measurement has been done or 
from the theoretical value of the total routhian surface (TRS) calculations. 
The intrinsic states within an energy window of 3.5 MeV around the Fermi surface are
considered. This gives rise to the size of the basis space, $|\phi_\kappa \ket$ 
in Eq. (1), of the order of 40. In the
second step, these basis states are projected to good angular-momentum states, and the
projected basis is then used to diagonalize the shell model Hamiltonian. 
The detailed calculations have been carried out for Kr-, Sr- and Zr-isotopes and the
results are discussed in the following subsections.
The band diagram \cite{hara91}, which gives the projected energies
for the configurations close to the Fermi surface is shown in Figs. 1,2 and 3
to explain the 
underlying physics. It should be noted that only a few of the most important configurations  
are plotted in the band diagrams, although many more are included in the calculations.

\subsection{Kr isotopes}


The proton-rich Kr-isotopes depict a variety of 
coexistence of prolate and oblate shapes in the low-spin region.
For these nuclei, the low-lying states have an oblate deformation
and the states above $I^\pi = 4^+$ are associated with a prolate shape. 

The first experimental information on transitions in $^{72}$Kr \cite {va97} was 
later confirmed in Refs. \cite {de90,sk98}. Recently, the level scheme has been 
extended up to $I^\pi = 16^+$ \cite {kr72}. The experimental value of the
transition strength is available only for $I^\pi = 8^+$. Following the indication in 
Ref. \cite {va97} that the yrast band at higher spins is prolate in nature, we
have calculated the $^{72}$Kr with a prolate basis deformation
of $\epsilon_2 = 0.36$. As shown in the band diagram in Fig. 1, the g-band is crossed by
a pair of proton 2-qp bands and a neutron 2-qp band at about $I = 10$.
This gives rise to a simultaneous alignment of neutron and proton pairs. The
measured energies and moment of inertia are compared with the calculated values in
Figs. 4 and 5. 
The calculated $Q_t$ values are shown in Fig. 6. 
The only data point available is in good agreement with our calculations.

Recent lifetime measurements in $^{74}$Kr up to $I^\pi = 18^+$ \cite {algora00} 
indicate a pronounced shape change induced by quasiparticle 
alignments. 
The $[440]\frac{1}{2}$, $[431]\frac{3}{2}$, $[422]\frac{5}{2}$ neutron orbitals and
the $[440]\frac{1}{2}$, $[431]\frac{3}{2}$, $[422]\frac{5}{2}$ proton orbitals play a major 
role in 
determining the nuclear shape at the high-spin region. 
In order to describe the 
high-spin states properly, we have used the prolate deformation of $\epsilon_2 $ = 0.37 in
the PSM calculations for $^{74}$Kr. 
The band diagram
is shown in
Fig. 1. The neutron 2-qp state $\nu g_{9/2} [3/2,5/2]$
and two proton 2-qp states $\pi g_{9/2}[1/2,3/2] $ and $\pi g_{9/2}[3/2,5/2] $ 
cross the g-band around $I = 14$. This corresponds to the 
alignment of neutron and proton pairs as observed in the experiment. 
The yrast band, consisting of the lowest energies after diagonalization at each spin, 
is plotted in Fig. 4 to compare with the available experimental data. 
These values are also displayed in Fig. 5 in a sensitive plot
of moment of inertia as a function of square of the rotational frequency.
For $^{74}$Kr, there is a very good agreement between the theory and 
experiment in both plots, except for the lowest spin states. 
In Fig. 6, the $Q_t$ values from the theory and experiment are 
compared. The sudden fall in the $Q_t$ at $I = 14$, which is described correctly 
by our calculations, is associated with the first
band crossing (see Fig. 1). Due to the occupation of the fully aligned orbitals after this 
band crossing, the $Q_t$ value becomes smaller for the higher spin states. 
Above $I = 18$, maximum contribution
to the yrast states 
comes from the configuration based on the 4-qp state $\nu g_{9/2} [3/2,5/2]$ + 
$\pi g_{9/2}[1/2,3/2] $. 

The energy levels and their lifetimes in $^{76}$Kr have been measured up to $I^\pi = 22^+$
along the yrast band \cite {gr89,gp99}. Lifetime measurements indicate a large
deformation in this nucleus. A simultaneous alignment of proton and neutron 
pairs is observed. For this nucleus, the calculations are performed 
for a prolate deformation of $\epsilon_2 $ = 0.36. 
The band diagram of $^{76}$Kr is shown in Fig. 1. 
The g-band 
is crossed by the neutron 2-qp band 
$\nu g_{9/2} [3/2,5/2]$ and two proton 2-qp bands 
$\pi g_{9/2}[1/2,3/2] $ and $\pi g_{9/2}[3/2,5/2]$ between $I = 12 - 14$. This 
corresponds to the simultaneous alignments of neutron and proton pairs. 
At higher spins, the wave function of the yrast states receives contribution from the
4-qp configurations. Though the major component of the wave function 
for the higher-spin states comes from the 4-qp state based on 
$ \nu g_{9/2} [3/2,5/2] + \pi g_{9/2}[1/2,3/2]$ as shown in the plot,    
we find some amount of $K$-mixing with other 4-qp states 
leading to a non-axial shape. The comparisions of the calculated
energies and $Q_t$ values with experimental data show
good agreement as shown in Figs. 4, 5 and 6.
As in the $^{74}$Kr case, the drop in $Q_t$ value at spin $I = 14$ is obtained 
due to the occupation of the aligned states.

Although good agreement between theory and experiment for the $^{72}$Kr spectrum is obtained in
Fig. 4, clear discrepancy can be found in the more sensitive plot of Fig. 5. 
Besides the problem in the description of the moment of inertia 
for the lowest spin states 
as in the $^{74}$Kr and $^{76}$Kr cases,  
the current calculation can not reproduce the fine details around the backbending region.
For this $N=Z$ nucleus, there has been an open question of whether the proton-neutron
pair correlation plays a role in the structure discussions. 
It has been shown that with the renormalized pairing interactions within 
the like-nucleons in an effective
Hamiltoinan, one can account for the $T=1$ part of the proton-neutron pairing \cite{frau99}. 
However, whether the renormalization is sufficient for the complex region
that exhibits the phenomenon of band crossings, 
in particular when both neutron and proton pair alignments 
occur at the same time is an interesting question to be investigated.

\subsection{Sr isotopes}

Light strontium isotopes are known to be among the well-deformed nuclei in the mass-80 
region. 
In contrast to the Kr isotopes, data for Sr, including the most recent ones \cite {wi00},
do not suggest a backbending or upbending in moment of inertia (see Fig. 5).  
For $^{76}$Sr, the only experimental work was reported in Ref. \cite {li901},
where the energy of the first excited state was identified. Because of
shell gap at $N = 38$, this nucleus is found to be highly deformed.
We have calculated this nucleus with the basis deformation 
of $\epsilon_2$ = 0.36. It is found in Fig. 2 that the proton 2-qp band $\pi g_{9/2}[3/2,5/2]$
and the neutron 2-qp band $\nu g_{9/2} [3/2,5/2]$ 
nearly coincide with each other for the entire low-spin region, and cross the g-band at $I = 14 $. 
Just above $I = 16$, the 4-qp band based on the configuration 
$\nu g_{9/2} [3/2,5/2] + \pi g_{9/2}[3/2,5/2] $ crosses the 2-qp bands and becomes the lowest
one for higher spins. However, both of the first and the second band crossings are  
gentle with very small crossing angles. Therefore, the structural change along the yrast band is
gradual, and the yrast band obtained as
the lowest band after band mixing seems to be unperturbed.  
This is true also for the isotopes $^{78}$Sr and $^{80}$Sr as we shall see below. 
Thus, although band crossings occur in both of the Kr and Sr isotopes, smaller crossing angle
results in a smooth change in the moments of inertia in Sr, and 
this picture is very different from what we have seen in the Kr band diagrams in Fig. 1,
where sharper band crossings lead to sudden structure changes. 

Ref. \cite {li82} reported the positive parity yrast cascade in $^{78}$Sr
up to $I^\pi = 10^+$. Their measurement of the lifetimes of the first two excited states 
indicate the $E2$ transition strengths of more than 100 Weisskopf units. Refs.  
\cite {gr891,da87} extended this sequence up to $I^\pi = 22^+$ . A broad band crossing
was observed around $I = 12$ and two bands in $^{78}$Sr seem to interact strongly
over a wide range of spin states. To study these effects, the band energies are
calculated for $^{78}$Sr, assuming a prolate deformation $\epsilon_2 $ = 0.36. 
As shown in the band diagram of $^{78}$Sr in Fig. 2, the band 
based on the proton 2-qp state occupying $[431]\frac{3}{2}$ and $[422]\frac{5}{2}$ orbitals
crosses the g-band at $I = 12$ at a very small angle. This implies a 
smooth structure change in the yrast band and explains the gradual 
alignment behavior as observed in the experiment. In fact, the neutron 2-qp
band based on $[431]\frac{3}{2}$ and $[422]\frac{5}{2}$ orbitals
contribute to the
yrast levels between $I = 14$ to 18. 
The calculated energies and moment of inertia 
are compared with the experimental data in Figs. 4 and 5.
In Fig. 6, the calculated $Q_t$ values show a smooth behavior with a slight 
decreasing trend for the entire spin region, which is 
in contrast to the sudden drop in the $Q_t$ curve  
at $I = 14$ in its isotone $^{74}$Kr.   
The current experimental data exist only for the lowest
two states. To test our prediction, the measurements for higher spins are desirable. 

In a recent experiment \cite {wi00}, $^{80}$Sr has been studied up to 
$I^\pi = 22^+$ along the yrast band. To observe the shape evolution along the
yrast band, the lifetime of levels up to $I = 14^+$ has been measured. 
At low spin, a prolate shape with $\beta_2 \approx 0.35 $ has been reported
\cite {wi00}. 
Taking $\epsilon_2 = 0.34$,
the band diagram of
$^{80}$Sr is calculated and displayed in Fig. 2. 
A proton 2-qp band based on $\pi g_{9/2}[3/2,5/2] $ crosses
the g-band between $I = 10$ and 12 and this corresponds to the first proton pair 
alignment. At higher spin values, the yrast states get contribution from a large 
number of 4-qp configurations with different $K$-values (These bands are not shown in Fig. 2). 
The calculated spectrum is compared with data in Figs. 4 and 5.
The agreement is satisfactory. In particular, the smooth evolution of 
moments of inertia is reproduced in Fig. 5.
Two sets of measured $Q_t$ values
\cite {wi00,da87} are compared with our calculations. 
The pronounced fluctuations in the $Q_t$ values from the early experiment \cite {da87}
have been removed by the new measurement in Ref. \cite {wi00}. 
However, our calculation does not reproduce the newly measured $Q_t$ at $I=14$.

\subsection{Zr isotopes}
The self-conjugate isotope $^{80}$Zr was studied up to $I^\pi = 4^+$ by the 
in-beam 
study \cite {li87}. The excitation energy of the 2$^+$ state indicates a
large deformation for this nucleus.  
In Fig. 3, the band diagrams of the proton-rich Zr isotopes are shown. The 
band diagram of $^{80}$Zr is calculated with $\epsilon_2$ = 0.36. 
For the positive parity band, the [431]3/2 and [422]5/2 quasiparticle states lie close to 
the Fermi levels of both neutrons and protons.
These orbitals play an important role in determining the alignment 
properties in $^{80}$Zr. Between $I = 12$ and 14, the proton and neutron 2-qp bands 
cross the ground band at the same spin. These 2-qp bands
consist of two $g_{9/2}$ quasiparticles in the [431]3/2 and [422]5/2 orbitals, 
coupled to $K = 1$. After this band crossing,
the yrast band gets maximum contribution from both of the proton 2-qp and 
neutron 2-qp states, as can be seen in Fig. 3. 
Immediately following the first band crossing, a 4-qp band crosses the 2-qp
bands at $I = 16$. In the plots of moment of inertia in Fig. 5, we see
that the total effect of the first and second band crossings produces a 
smooth upbending. In the $Q_t$ plots in Fig. 6, a two-step drop in the $Q_t$ values
is predicted, which is due to the two successive band crossings.  

Let us now look at the $N = Z + 2$ isotope of Zr. Recently, $^{82}$Zr has been studied
up to $I^\pi = 24^+$ \cite {ru97}. Lifetime measurements up to $I^\pi = 14^+$ along 
the yrast line \cite {paul97,ru97} indicate a 
smaller deformation $\epsilon_2 $ = 0.29 for this nuclei. The band diagram
is shown in Fig. 3. Here, the 2-qp band consists of two $g_{9/2}$ 
quasi-protons in [431]3/2 and [422]5/2 orbitals 
crosses the g-band at $I = 10$. Beyond that spin, this 2-qp proton 
configuration contributes maximum to the yrast states up to $I = 16$, where
a 4-qp band crosses the proton 2-qp band. This 4-qp configuration is made from
two quasi-protons in [431]3/2 and [422]5/2 orbitals and two
quasi-neutrons in [431]3/2 and [422]5/2 orbitals.
These two separate band crossings 
were observed in the experiment \cite {paul97,ru97}.
As can be seen from Figs. 4 and 5, a remarkable agreement between our calculation
and experimental data are obtained. 
The observed complex behavior of the two-step upbending in moment of inertia
in $^{82}$Zr is correctly reproduced.  
As shown in Fig. 6, for $^{82}$Zr, our theory predicts a small kink at $I = 10$ 
in the $Q_t$ value corresponding to the band crossing of the 
proton 2-qp band. At present the accuracy of the experimental values of $Q_t$ \cite 
{paul97,ru97} is not enough to confirm this kink. Up to $I = 12$, the 
measured transition
quadrupole moments match quite well with the calculations. At spin $I = 16$,
the dip in the calculated $Q_t$ value is because of the second band 
crossing. A close look at the components of the wave functions of the yrast
states, shows the mixing of different intrinsic states increases after
the first band crossing around $I = 10$. Thus this nucleus loses the
axial symmetry at higher spin states. 

For the $N = Z+4$ isotope of Zr, the deformation is less than the lighter ones 
\cite {sch96}. 
The band diagram shown in Fig. 3 is based on calculations 
with the basis deformation of $\epsilon_2 = 0.22 $.
At least two proton 2-qp bands based on 
$\pi g_{9/2}[3/2,5/2] $ and $\pi g_{9/2}[1/2,5/2] $ configurations 
cross the 
g-band at $I = 8$, corresponding to the proton alignment. 
After the first band crossing till $I = 14$, these two 2-qp bands lie together 
and continue to interact with each other. Then one 
finds a neutron alignment due to the lowering of a 4-qp band 
between spin $I = 14$ and 16. Starting from $I = 18$, two 4-qp bands nearly coincide 
for the higher spin states. The calculated energies of the yrast band 
are compared with that of the adopted
level scheme \cite {sch96} in Figs. 4 and 5. 
In Fig. 5, one can see that the interesting pattern of experimental moment of
inertia in $^{84}$Zr has been qualitatively reproduced although our calculation
exaggerates the variations in the data. 
To see the shape evolution up to high spins, the
experimental transition quadrupole moments are compared with the calculated
values in Fig. 6.
The measured $Q_t$ values for $^{84}$Zr \cite {sch96} show different
behavior from those of the neighboring even-even nuclei with an increasing trend 
after the second band crossing. 
Though the calculated values of 
$Q_t$ at higher spins are lower than the experimental data, the qualitative 
variation is well reproduced.

The only experiment that measured g-factors has recently been performed for $^{84}$Zr \cite{te99}.
Though, there are big uncertainties, the three data points suggest the rigid rotor value $Z/A$.  
As we can see in the theoretical diagram given in Fig 7, 
g-factors for the low spin states in this nucleus are predicted to have a rigid body character 
with a nearly constant value up to $I = 6$. 
However, a more than doubled value is predicted for $I = 8$ with a decreasing trend thereafter.  
This pronounced jump in g-factor is obviously due to the proton 2-qp band crossings at $I = 8$,
as discussed above (see also Fig. 3). The wave functions of yrast states at $I = 8$ to 14 
are dominated by the proton 2-qp states, thus enhancing the g-factor values.   
Therefore, measurement extended to higher spins in this nucleus will be a strong test
for our predictions. Similar g-factor increasing at the first band crossing should also
appear in the neighboring nuclei, for example in $^{82}$Zr, where proton 2-qp bands
alone dominate the yrast states.

\section{Conclusion}

The study of the proton-rich mass-80 nuclei is not only interesting from the structure
point of view, but it also has important implications in the 
nuclear astrophysical
study \cite{rp98}.  
In this paper, we have for the first time performed a systematic study 
for the yrast bands of $T_z$ = 0, 1 and 2 isotopes of 
Kr, Sr and Zr within the framework of the projected shell model approach. 
We have employed the quadrupole plus monopole and quadrupole pairing force in the 
Hamiltonian, and the major shells with N = 2, 3 and 4 have been included for the
configuration space. 

The proton-rich mass-80 nuclei exhibit many phenomena that are quite unique
to this mass region. The structure changes are quite pronounced among the
neighboring nuclei and can be best seen in the sensitive plot of Fig. 5.   
For the Kr nuclei, a clear backbending is observed for all the three isotopes,
while for the Sr isotopes no backbending is seen. 
However, the Zr nuclei exhibit both the first and the second upbends. 
The transition quadrupole moments show corresponding variations. 
For these variations, 
we have obtained an overall qualitative and in many cases quantitative
description. 
These variations can be understood by the mixing of various configurations 
of the projected deformed
Nilsson states.
In particular, they are often related to band crossing phenomenon.

Despite the success mentined above, our calculations fail to reproduce states at very low spins, 
as can be
clearly seen in the moment of inertia plot of Fig. 5 and the $Q_t$ plot of Fig. 6.
The present model space which is constructed from the intrinsic 
states with a fixed
axial deformation may be too crude for the spin region  
characterized by shape coexistence. 
To correctly describe the low-spin region, one needs to enrich the shell model basis.
Introduction of triaxiality in the deformed basis combined with three dimensional
angular momentum projection \cite{sheikh99} certainly spans a richer space. 
However, in order to study the high-spin states discussed in this paper, 
the current space of the triaxial PSM \cite{sheikh99} must be extended  
by including multi-quasiparticle states. 
The other possibility is to perform an investigation with the generator coordinate method 
\cite{hara99} which takes explicitly the shape evolution into account. 

\begin{figure}
\leavevmode \hspace{-0.0cm}\hbox{\epsfxsize = 8.0 cm \epsfysize = 8.0 cm
\epsffile{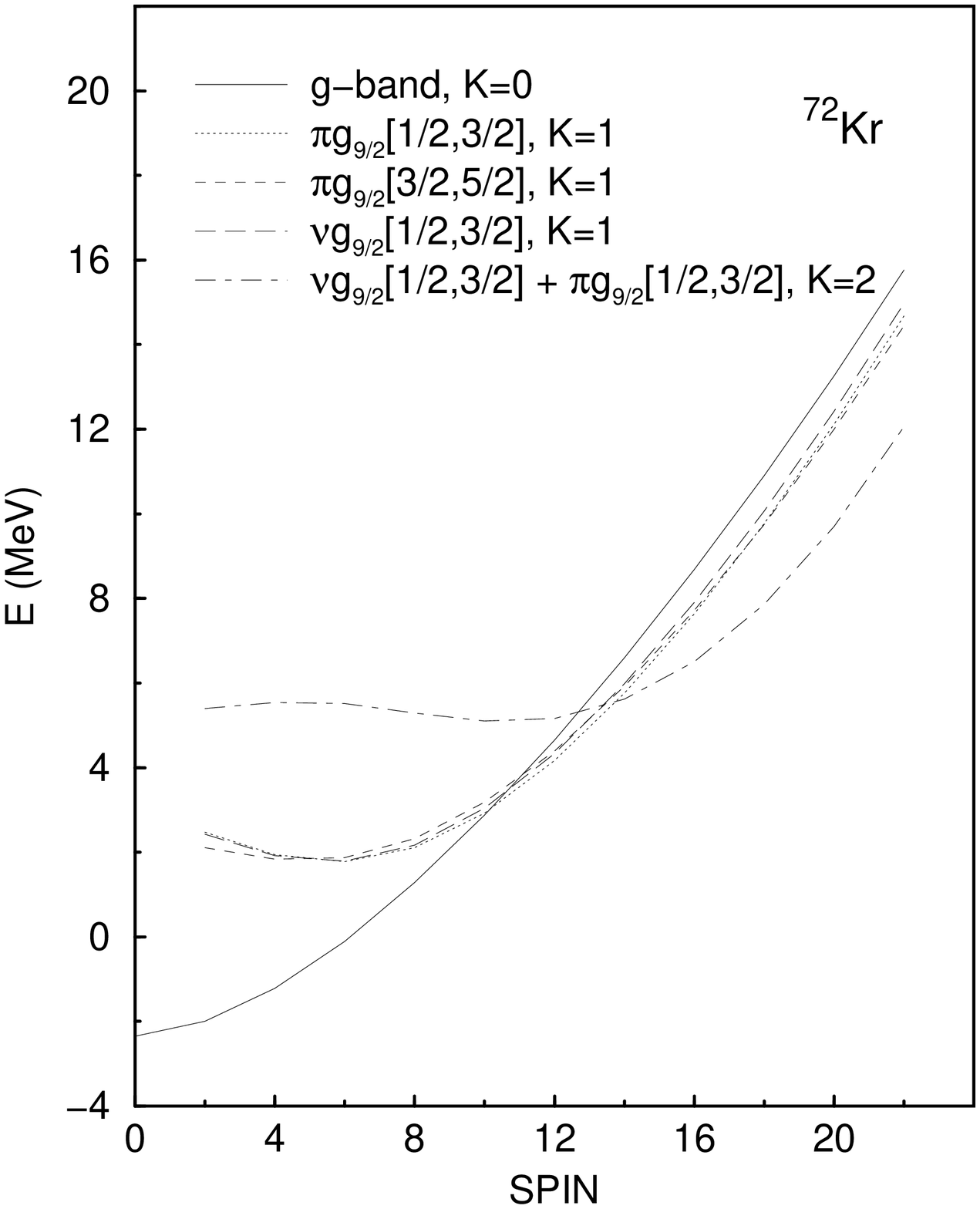}}
\vspace{-0.5cm}
\leavevmode \hspace{-0.0cm}\hbox{\epsfxsize = 8.0 cm \epsfysize = 8.0 cm
\epsffile{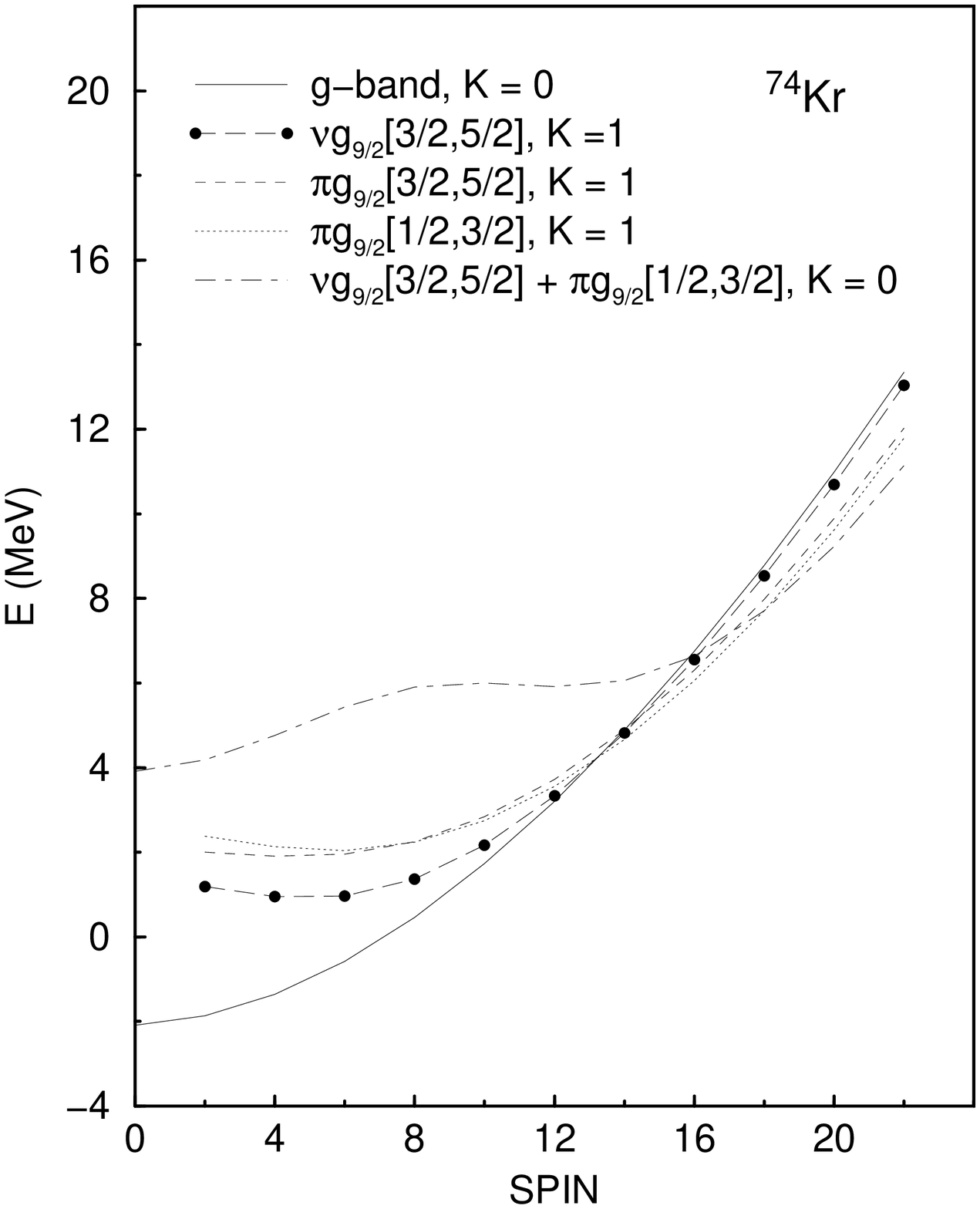}}
\vspace{-0.5cm}
\leavevmode \hspace{1.8cm}\hbox{\epsfxsize = 8.0 cm \epsfysize = 8.0 cm
\epsffile{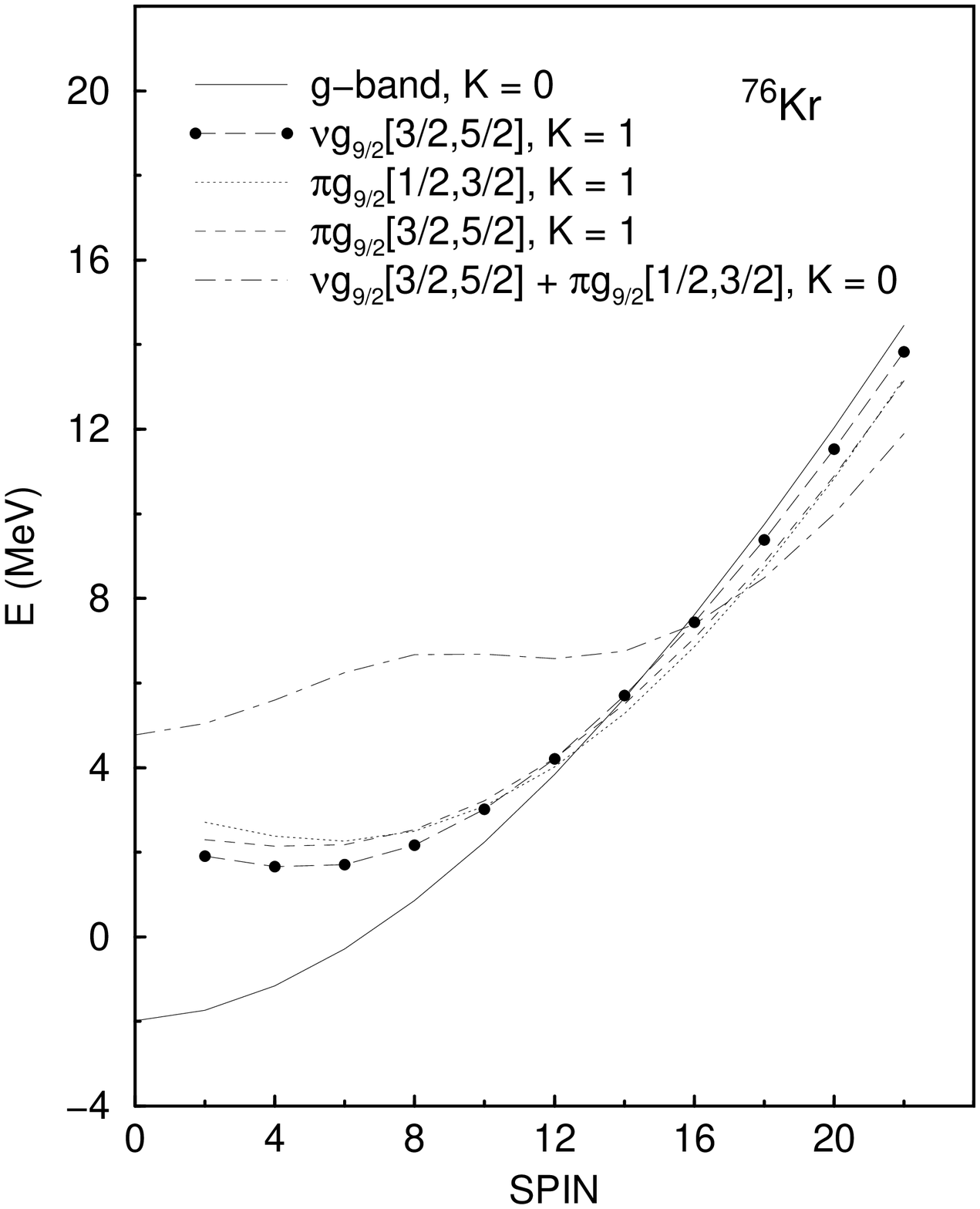}}
\caption{Band diagrams for Kr isotopes. }
\end{figure}

\begin{figure}
\leavevmode \hspace{-0.0cm}\hbox{\epsfxsize = 8.0 cm \epsfysize = 8.0 cm
\epsffile{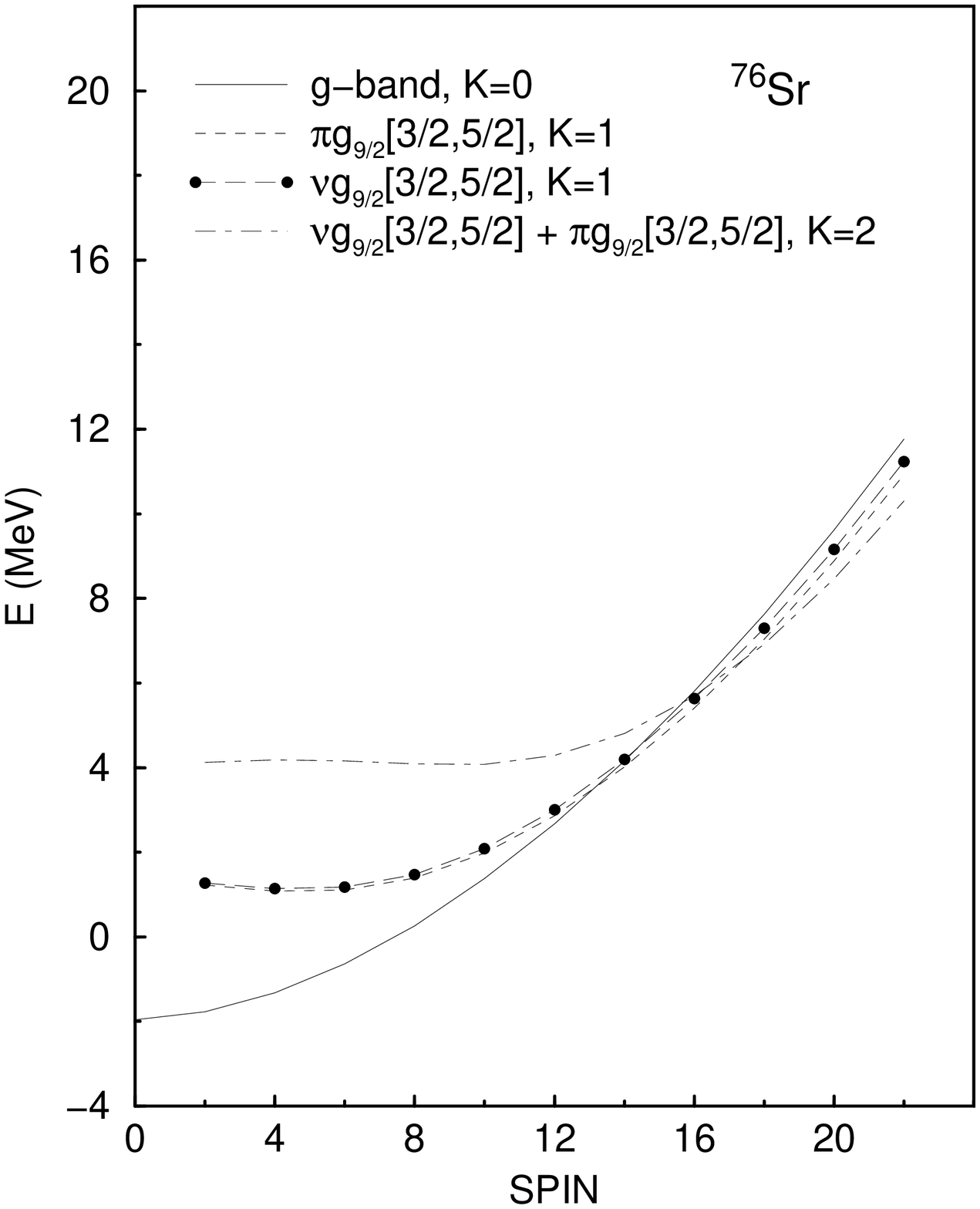}}
\vspace{-0.5cm}
\leavevmode \hspace{-0.0cm}\hbox{\epsfxsize = 8.0 cm \epsfysize = 8.0 cm
\epsffile{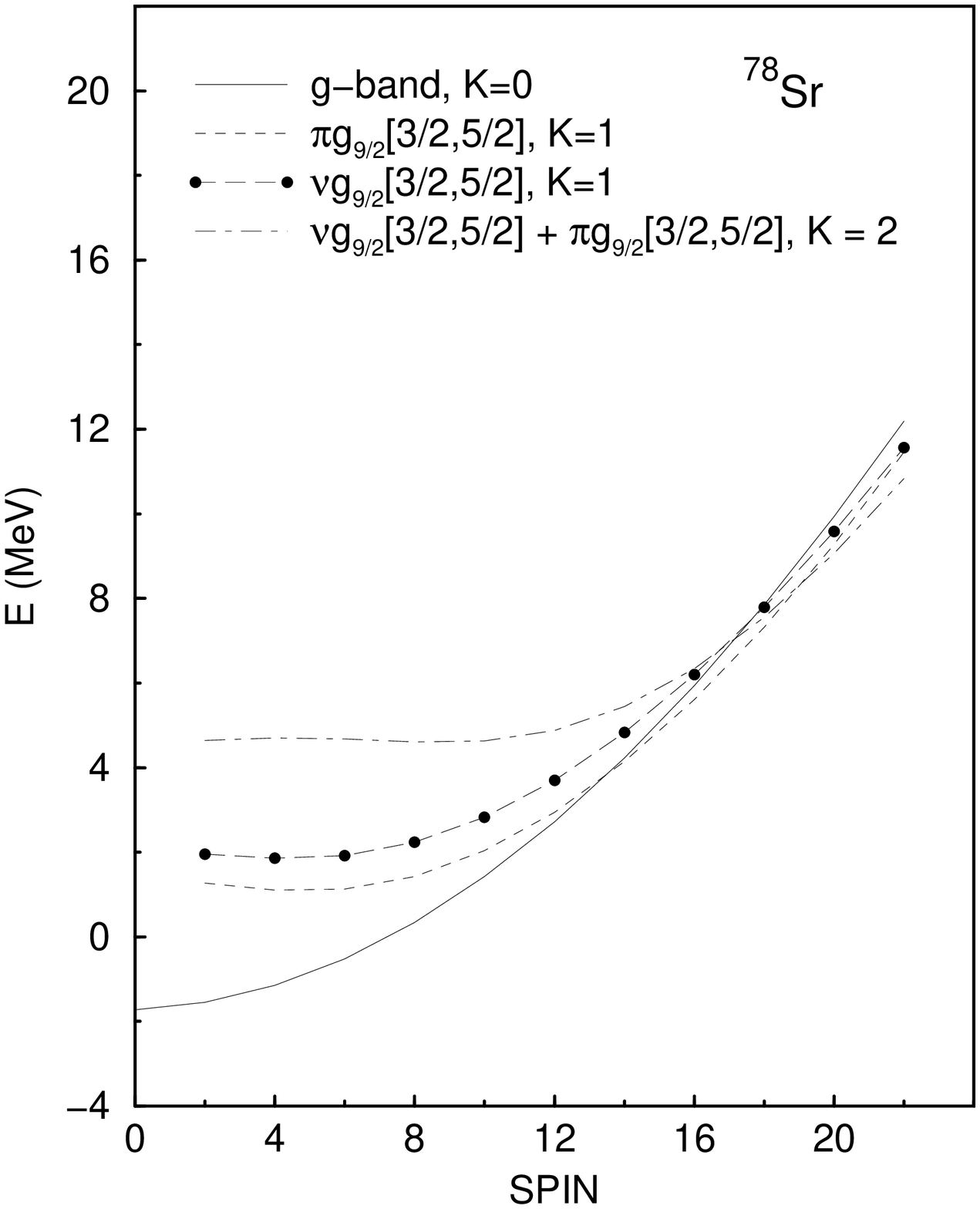}}
\vspace{-0.5cm}
\leavevmode \hspace{1.8cm}\hbox{\epsfxsize = 8.0 cm \epsfysize = 8.0 cm
\epsffile{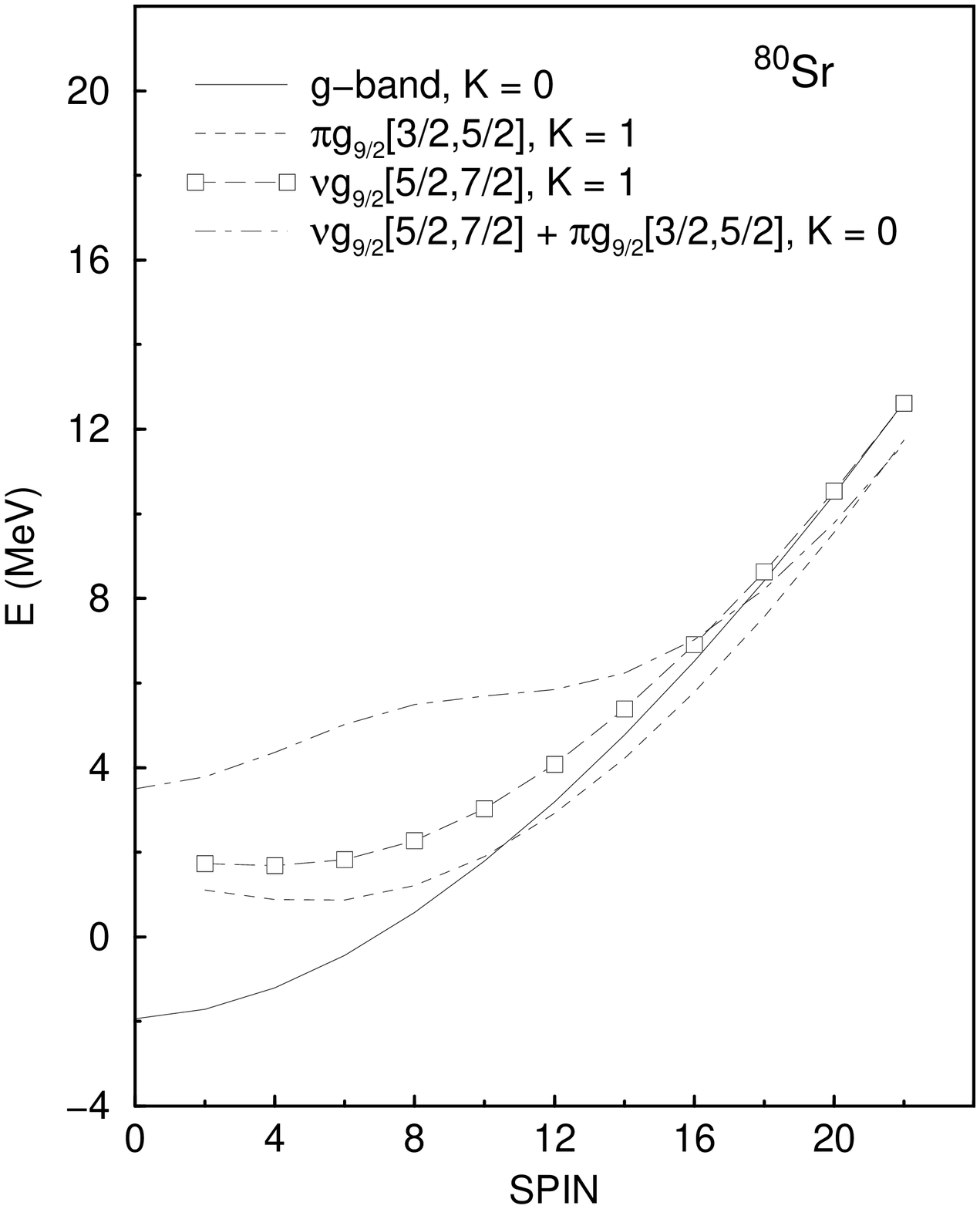}}
\caption{Band diagrams for Sr isotopes. }
\end{figure}
\begin{figure}
\leavevmode \hspace{-0.0cm}\hbox{\epsfxsize = 8.0 cm \epsfysize = 8.0 cm
\epsffile{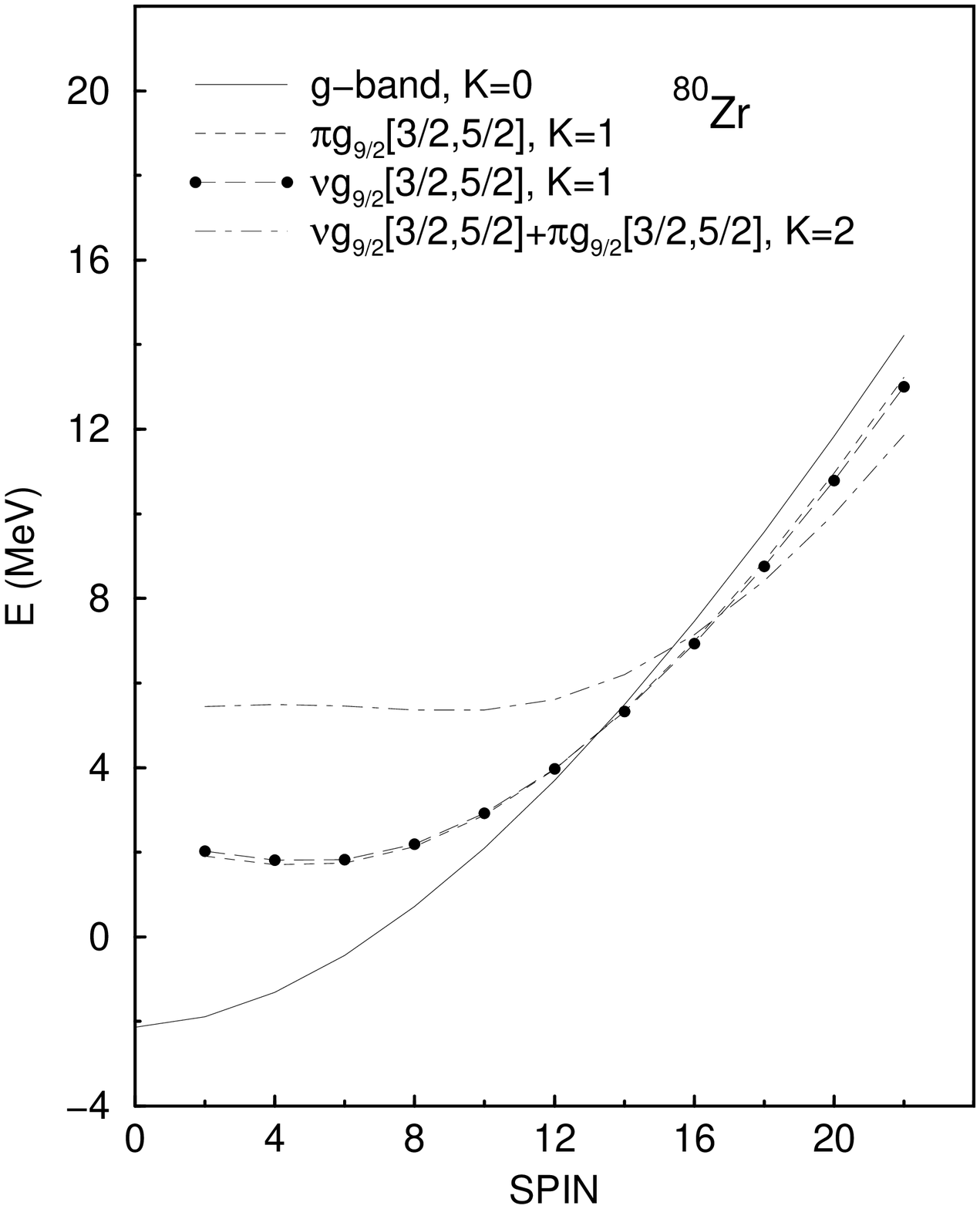}}
\vspace{-0.5cm}
\leavevmode \hspace{-0.0cm}\hbox{\epsfxsize = 8.0 cm \epsfysize = 8.0 cm
\epsffile{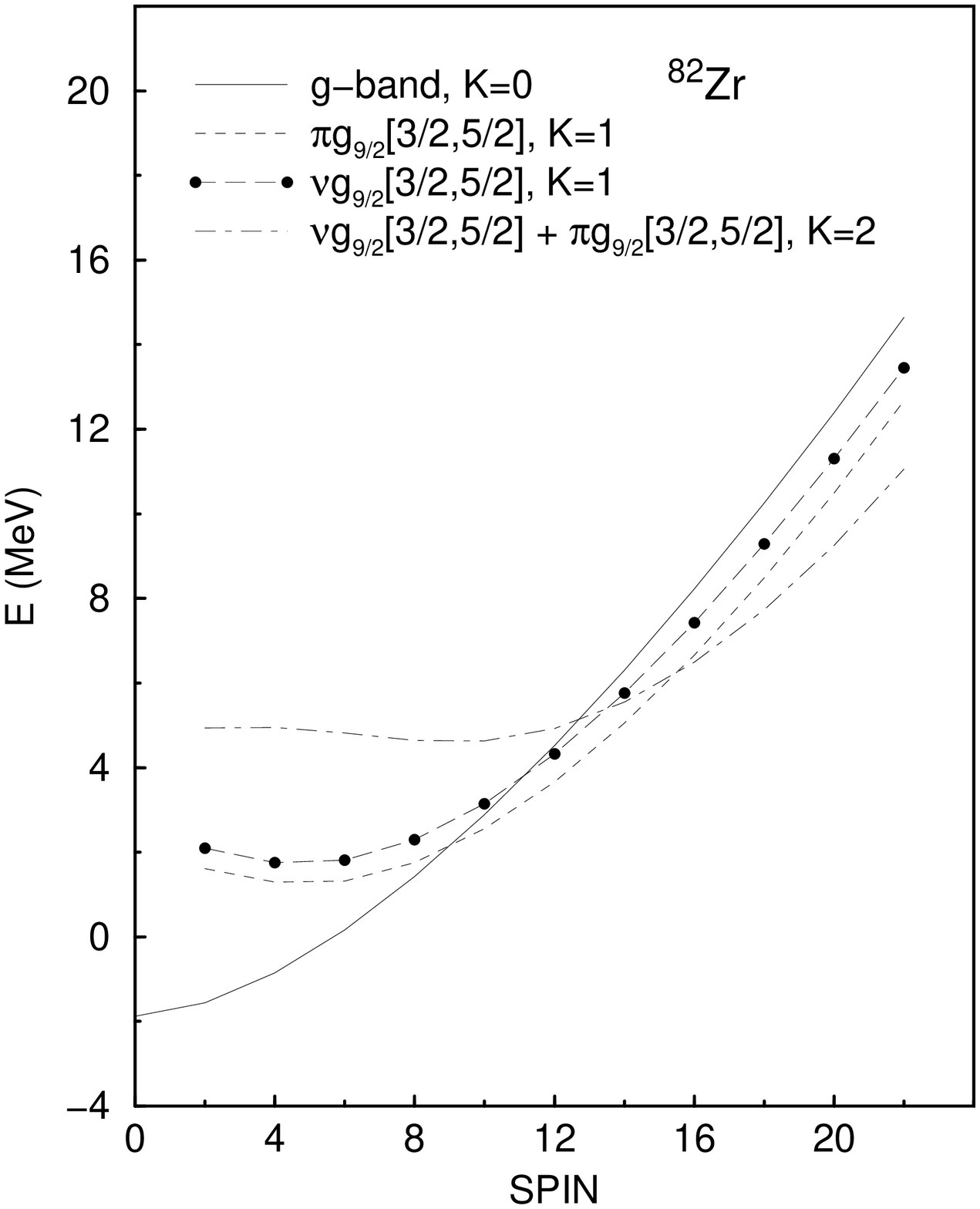}}
\vspace{-0.5cm}
\leavevmode \hspace{1.8cm}\hbox{\epsfxsize = 8.0 cm \epsfysize = 8.0 cm
\epsffile{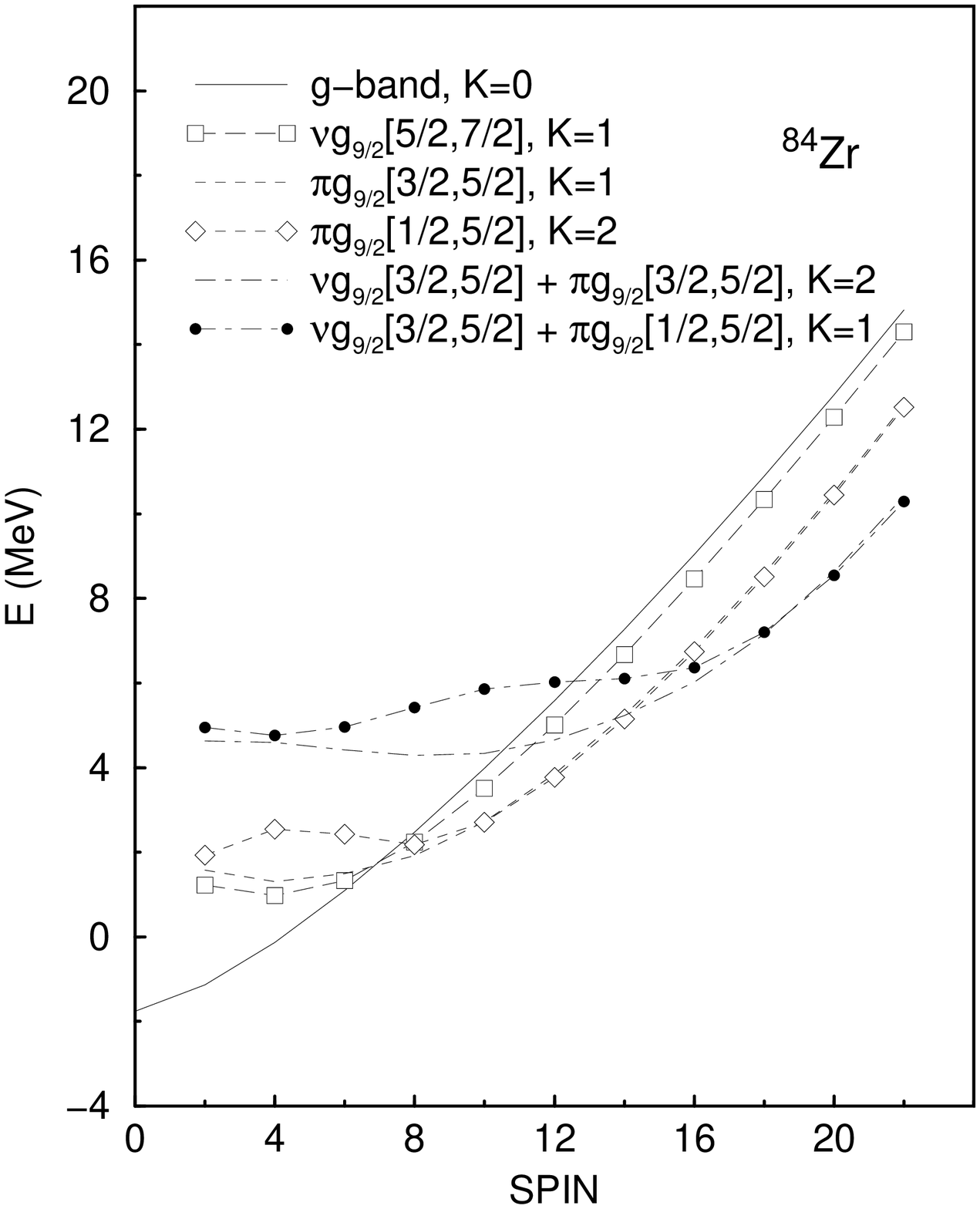}}
\caption{Band diagrams for Zr isotopes. }
\end{figure}

\newpage
~
\vspace{1.5cm}
\begin{figure}
~
\vspace{1.5cm}
\leavevmode \hspace{-0.0cm}\hbox{\epsfxsize = 12.0 cm \epsfysize = 12.0 cm
\epsffile{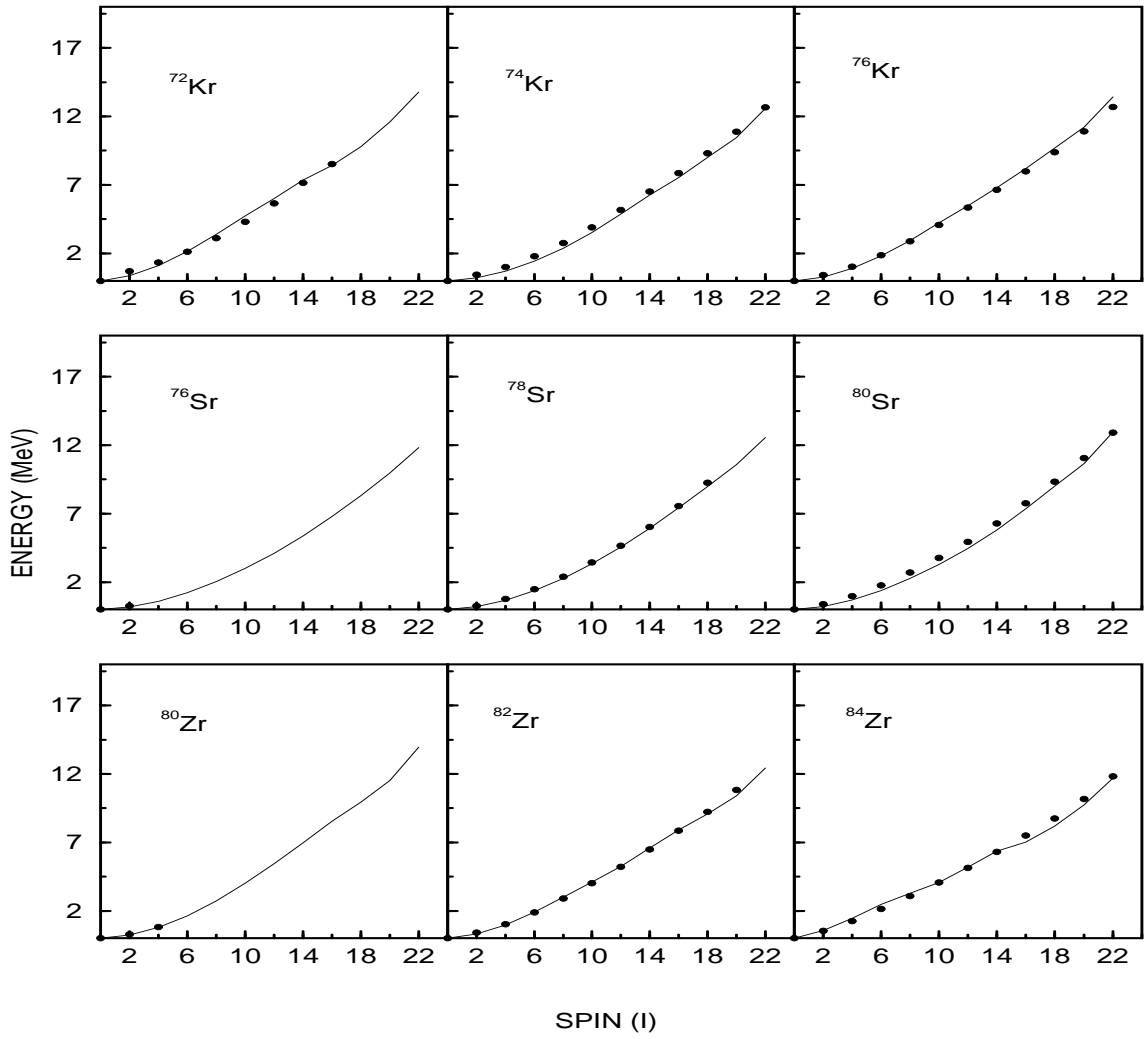}}
\caption{Comparison of the calculated energies $E(I)$ of the yrast bands with experimental data
for Kr, Sr and Zr isotopes.
The calculated yrast bands consist of the lowest states after diagonalization
at each angular
momentum $I$. }
\end{figure}

\newpage
~
\vspace{1.5cm}
\begin{figure}
~
\vspace{1.5cm}
\leavevmode \hspace{-0.0cm}\hbox{\epsfxsize = 12.0 cm \epsfysize = 12.0 cm
\epsffile{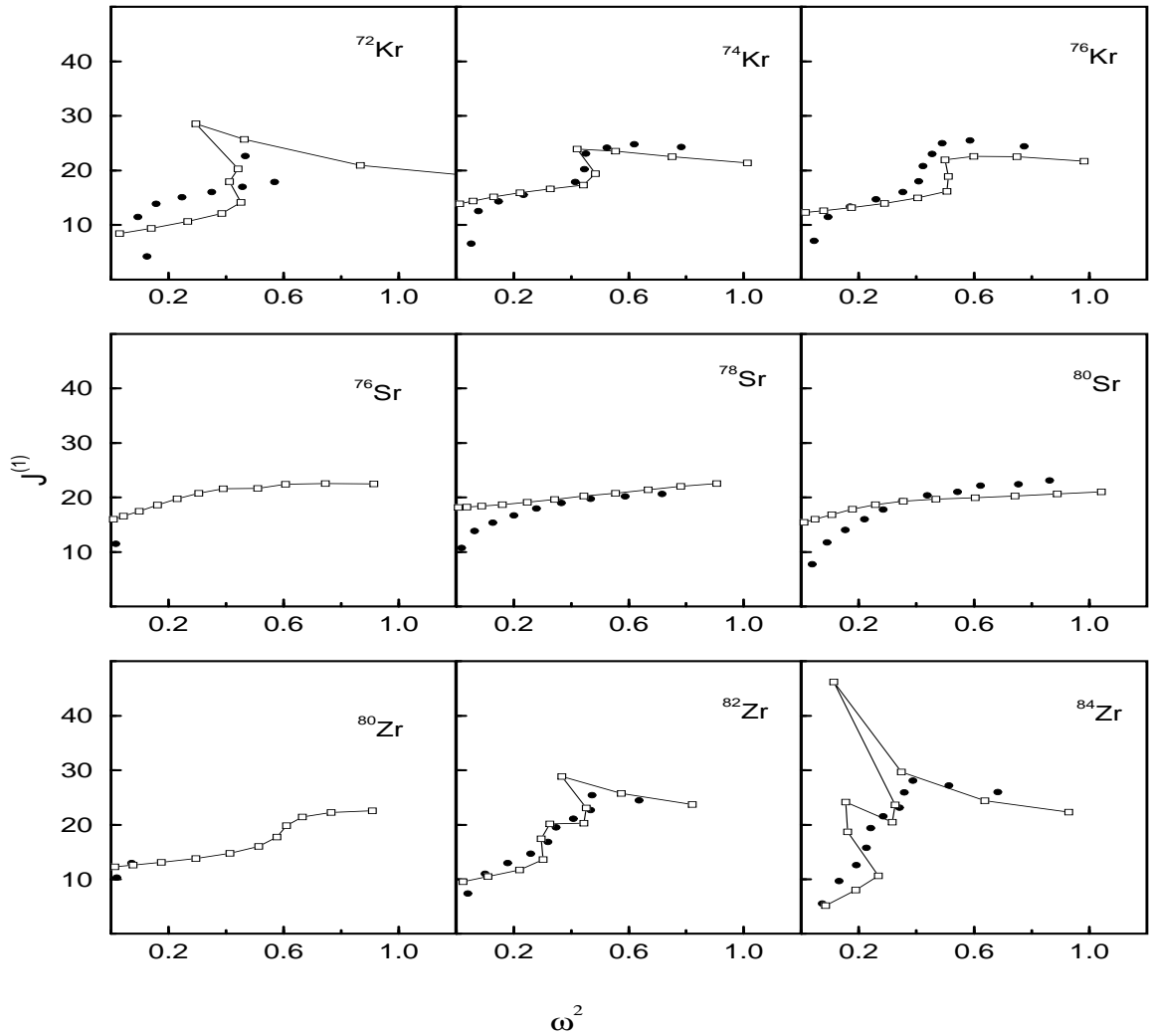}}
\caption{Comparison of calculated moments of inertia $J^{(1)}$ with experimental data as a
function of square of rotational frequency $\omega^2$ 
for Kr, Sr and Zr isotopes. These quantities are defined as
$J^{(1)} = (I - {1\over 2}) / \omega$ and $\omega = [E(I)-E(I-2)] / 2$ .
}
\end{figure}

\newpage
~
\vspace{1.5cm}
\begin{figure}
~
\vspace{1.5cm}
\leavevmode \hspace{-0.0cm}\hbox{\epsfxsize = 12.0 cm \epsfysize = 12.0 cm
\epsffile{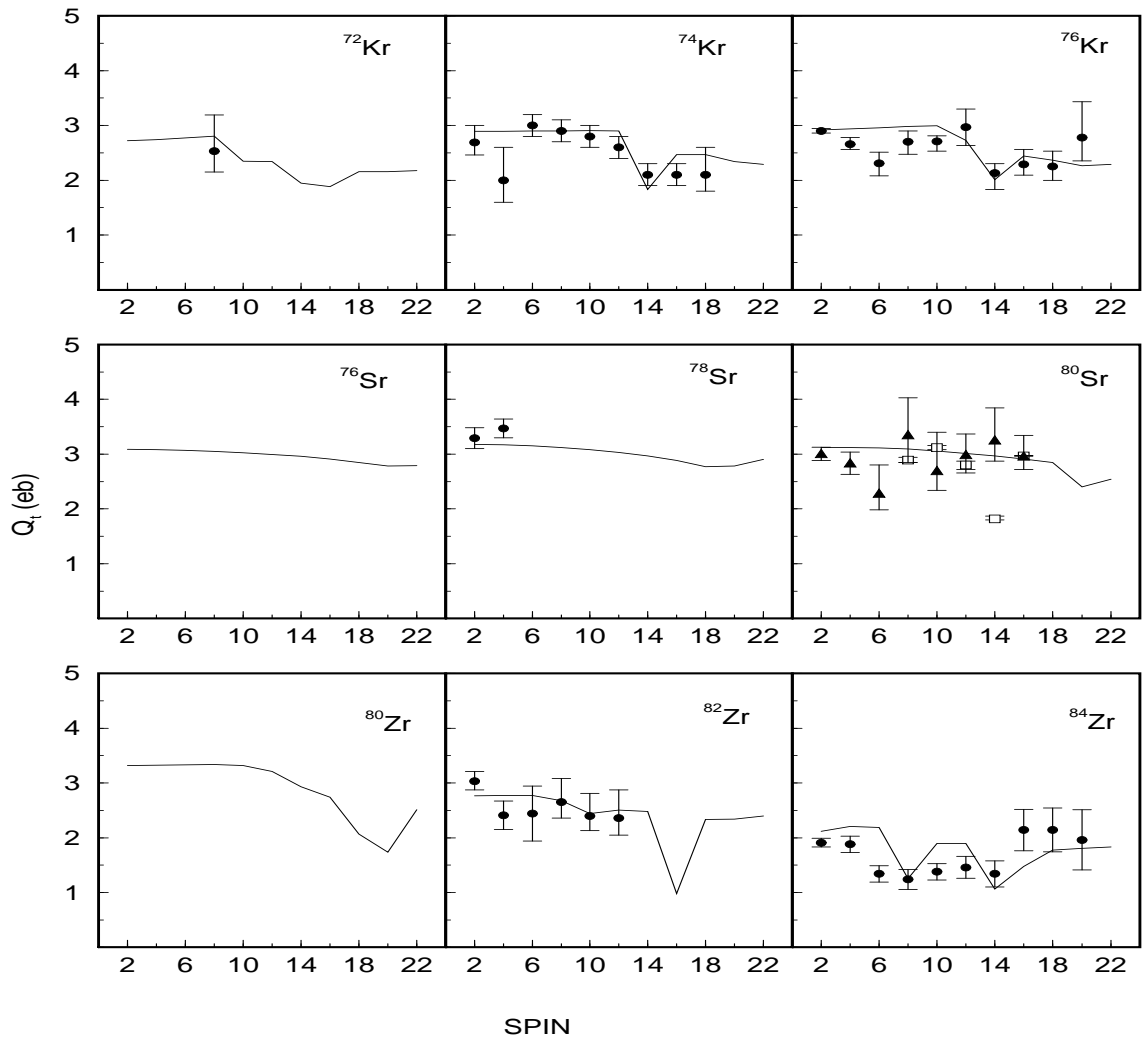}}
\caption{Comparison of calculated transitional quadrupole moments $Q_t$ with 
experimental 
values for Kr, Sr and Zr isotopes.}
\end{figure}

\newpage
~
\vspace{1.5cm}
\begin{figure}
~
\vspace{1.5cm}
\leavevmode \hspace{-0.0cm}\hbox{\epsfxsize = 12.0 cm \epsfysize = 12.0 cm
\epsffile{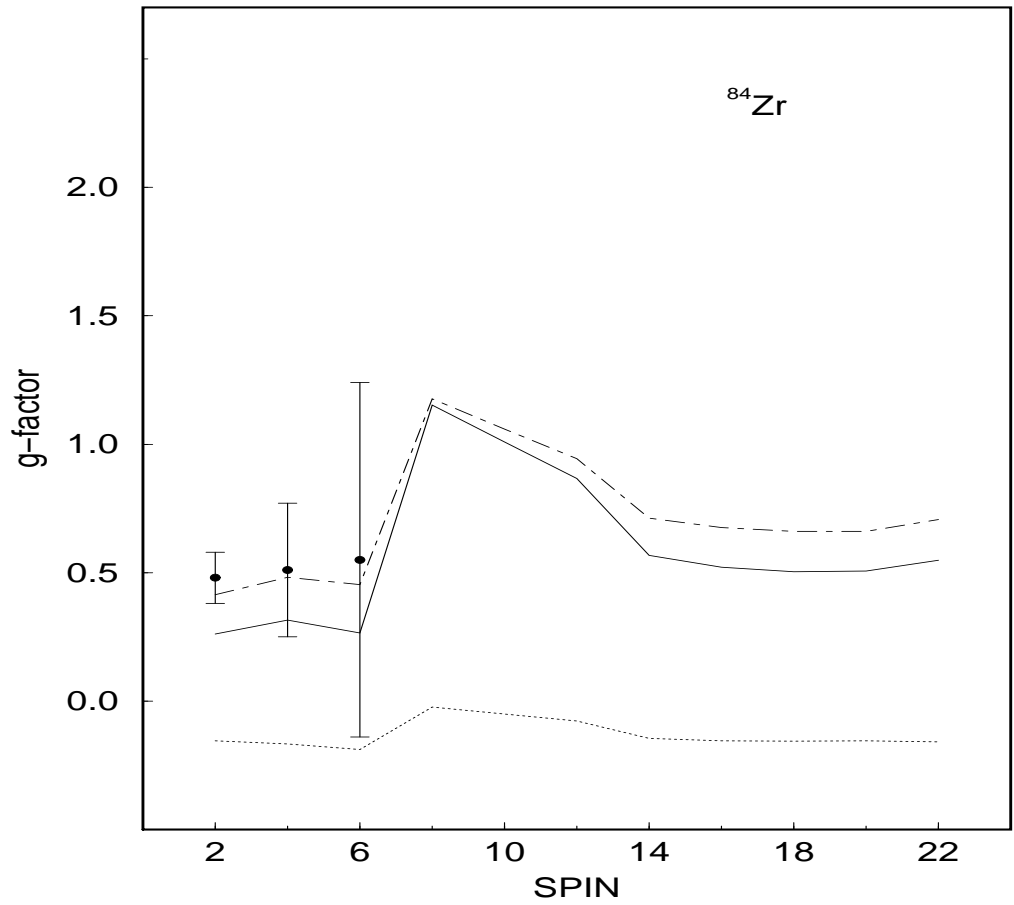}}
\caption{Comparison of calculated gyromagnetic factors 
with experimental values for $^{84}$Zr. For the theoretical values,
dashed (dotted) line represents the $g_\pi$ ($g_\nu$) part as given in Eq. (11),
and solid line gives the total g-value. }
\end{figure}

\end{document}